\documentclass[pre,aps,manuscript]{revtex4}
\usepackage{graphicx}
\usepackage{dcolumn}
\usepackage{bm}

\newcommand{\bs}{\bigskip}
\newcommand{\bc}{\begin{center}}
\newcommand{\ec}{\end{center}}

\begin{document}

\title{Non--Markovian response of complex quantum systems}
\author{ S.V.\surname{Radionov}
\footnote{Electronic address: sergey.radionov18@gmail.com}}
\affiliation{\textit{Institute for Nuclear Research, 03680 Kiev, Ukraine} }
\date{\today}

\begin{abstract}
We study the perturbative response of a complex quantum system 
on time changes of an external parameter $X$. The driven dynamics
is treated in adiabatic basis of the system's Hamiltonian $\hat{H}[X]$.
Within a random matrix approach we obtained non--Markovian Fokker--Planck
equation for the occupancy of given adiabatic state. We observed
normal diffusion regime of the driven quantum dynamics at
quite small values of the memory time defined by the time
scales of the $X$--correlations and energy--distribution
of the coupling matrix elements $(\partial \hat{H}/\partial X)_{nm}$.
Here the normal energy diffusion was found to drop out with
the width of the matrix elements' energy--distribution and
the diffusion may be significantly suppressed with the
decrease of the correlations between the matrix elements.
In the opposite limit of relatively large memory times we
obtained ballistic regime of the dynamics.

\end{abstract}

\maketitle


\section{Introduction}

In the paper we study the response of complex quantum systems
on an external parametric driving. Such a study can be important
to clarify the physics of dissipation appearing in dynamics
of macroscopic coordinates coupled to fast intrinsic degrees of 
freedom of complex many body systems. 
Usually dynamics of the complex systems is characterized by the 
absence of constants of motion (symmetricies) except trivial like 
the total energy and angular momentum. The spectra of atomic nuclei, 
quantum dots, mesoscopic systems and other systems show universal 
statistical properties which can be well modelled by random matrix 
ensembles. 

The first, who applied the random matrix approach to the description
of complex systems, were Gorkov and Eliashberg~\cite{ge}. They considered 
the absorption of photons by small metallic particles and found that the 
susceptibility of the system may show different dependence on temperature
for different random matrix ensembles of levels using to model the system's
spectrum. The problem of susceptibility of quantum systems to perturbations
has been developed further by many authors, see, for example~\cite{es,coh1,coh2}. 
Th other branch of interest in applying the random matrix approach 
is the study of quantum dissipation problem. Thus, Wilkinson in a
series of papers~\cite{wil1,wil2,wil3} discusses the rate of change 
of energy of the driven system in context of Landau--Zener transitions 
between levels. The same problem of the dissipation properties of
of many body systems is investigated in Refs.~\cite{kol,bul,ra,kar}.  
The main aim of the present investigation is to study different regimes
of driven dynamics of complex quantum systems within the random matrix
approach. We wish to see how the intrinsic properties of the system
may show up in its response on the external perturbation.   

The plan of the paper is as follows. In Sect.~\ref{dqd} we start from the
time--dependent Schr\"{o}dinger equation and introduce adiabatic basis
of the system's Hamiltonian. In the weak--coupling limit we get a
closed set of equations for the occupancies of adiabatic states. Then,
we apply the random matrix model and reduce the driven quantum dynamics
to non--Markovian Fokker--Planck equation. Different regimes of the
dynamics as a function of the parameters of the model are discussed in
Sect.~\ref{dr}. Finally, conclusions and discussion of the main results
 of the paper are given in the Summary.

\section{Driven quantum dynamics}

\label{dqd}

We start from the time--dependent Shr\"{o}dinger equation for the
time evolution of complex quantum system $H[X]$ driven by a single
external parameter $X(t)$
\begin{equation}
i\hbar\frac{\partial \Psi}{\partial t}=\hat{H}\Psi.
\label{schr}
\end{equation}
We also introduce an adiabatic basis of the system
\begin{equation}
\hat{H}\mu_n=E_n\mu_n,
\label{adiab}
\end{equation}
where adiabatic eigenfunctions $\mu_n[X]$ and eigenenergies $E_n[X]$ 
of the system's Hamiltonian are determined for each fixed value of the 
parameter $X$. Let us use the following expansion for the total wave 
function
\begin{equation}
\Psi(t)=\sum_n a_n(t)e^{i\phi_n(t)}\mu_n(X[t]),
\label{at}
\end{equation}
where quantum--mechanical phases $\phi_n$ are given by
\begin{equation}
\phi_n=\frac{1}{\hbar}\int_0^t E_n(X[t'])dt'.
\label{phi}
\end{equation}
Substituting Eqs.~(\ref{adiab}) and (\ref{at}) into 
Eq.~(\ref{schr}), we obtain an equation for the amplitudes 
$a_n(t)$
\begin{equation}
\frac{da_n}{dt}=-\dot{X}\sum_{m\neq n}\frac{{\cal M}_{nm}}{E_n-E_m}
e^{i(\phi_n-\phi_m)}a_m,
\label{anam}
\end{equation}
with matrix elements
\begin{equation}
{\cal M}_{nm}=\langle \mu_n|\partial\hat{H}/\partial X|\mu_m\rangle.
\label{Mnm}
\end{equation}
From Eq.~(\ref{anam}), a set of coupled equations defining how the 
occupancies of adiabatic states $|a_n|^2$ evolve with time is obtained,
\begin{equation}
\frac{d(|a_n|^2)}{dt}=-\dot{X}\left(\sum_{k\neq n}\frac{{\cal M}_{nk}}{E_n-E_k}
e^{i(\phi_n-\phi_k)}a_k a_n^*-\sum_{l\neq n}
\frac{{\cal M}_{nk}^*}{E_n-E_l}e^{-i(\phi_n-\phi_l)}a_l^* a_n \right),
\label{an2}
\end{equation}
\begin{eqnarray}
\frac{d(a_p a_n^*)}{dt}=\dot{X}\bigg(\frac{{\cal M}_{pn}}{E_n-E_p}
e^{i(\phi_p-\phi_n)}[|a_p|^2-|a_n|^2]-
\sum_{r\neq n,p}\frac{{\cal M}_{pr}}{E_p-E_r}
e^{i(\phi_p-\phi_r)}a_r a_n^*
\nonumber\\
-\sum_{s\neq {n,p}}
\frac{{\cal M}_{ns}^*}{E_n-E_s}e^{-i(\phi_n-\phi_s)}a_s^* a_n \bigg),~~~~~p\neq n
\label{apan}
\end{eqnarray}
where star denotes the complex conjugation. The initial conditions for
the system of equations (\ref{an2})--(\ref{apan}) are chosen such that 
initially only one given eigenstate $n$ is occupied,
\begin{equation}
(a_p a_n^*)(t=0)=\delta_{pn}.
\label{initial}
\end{equation}

Formally, one can obtain a closed equation for the occupancies $|a_n|^2$
themselves just by integrating over time both sides of Eq.~(\ref{apan}),
\begin{eqnarray}
(a_p a_n^*)(t)=\int_0^t dt'\dot{X}\bigg( \frac{{\cal M}_{pn}}{E_n-E_p}
e^{i(\phi_p-\phi_n)}[|a_p|^2-|a_n|^2]-
\sum_{r\neq n,p}\frac{{\cal M}_{pr}}{E_p-E_r}
e^{i(\phi_p-\phi_r)}a_r a_n^*
\nonumber\\
-\sum_{s\neq n,p}
\frac{{\cal M}_{ns}^*}{E_n-E_s}e^{-i(\phi_n-\phi_s)}a_s^* a_n \bigg),
\label{formal}
\end{eqnarray}
and subsequent substitution of the interference terms like $(a_p a_n^*)$ 
(\ref{formal}) into Eq.~(\ref{an2}). In this way, the right--hand side 
(rhs) of Eq.~(\ref{an2}) get a form of the perturbative expansion in 
terms of a parameter
\begin{equation}
\alpha=2\dot{X}(t)\sum_{k\neq n}\int_0^t dt'\dot{X}(t')
Re\left(\frac{{\cal M}_{nk}(t){\cal M}_{kn}(t')}{(E_n-E_k)(t)(E_n-E_k)(t')}
e^{i(\phi_n-\phi_k)(t)(\phi_k-\phi_n)(t')}\right)
\label{alpha}
\end{equation}
We consider the perturbation parameter $\alpha$ to be sufficiently small 
such that we are able to restrict ourselves by keeping only the lowest 
order terms in $\alpha$ in the right--hand side of Eq.~(\ref{an2}). 
Thus, we have
\begin{eqnarray}
\frac{d(|a_n|^2)}{dt}=2\dot{X}(t) \sum_{k\neq n}\int_0^t dt' \dot{X}(t')
[|a_k|^2-|a_n|^2](t')
\nonumber\\
Re\left(\frac{{\cal M}_{nk}(X[t]){\cal M}_{kn}(X[t'])}{(E_n-E_k)(t)(E_n-E_k)(t')}
e^{i(\phi_n-\phi_k)(t)(\phi_k-\phi_n)(t')}\right).
\label{an2ap}
\end{eqnarray}
Equation~(\ref{an2ap}) is an integro--differential equation determining
the time variations of the occupancy of the given quantum state $n$ due
to the interlevel transitions from all other states $k$. 

At this place, we apply the formalism of random matrix theory (RMT) 
and average the rhs of Eq.~(\ref{an2ap}) over suitably chosen statistics 
of randomly distributed energy spacings $E_n-E_k$ and off--diagonal matrix 
elements ${\cal M}_{nk}$. It is assumed that such an ensemble averaging 
can be performed independently over the spacings and matrix elements. 
First, energy spacings part of the ensemble averaging is defined as
\cite{ge}
\begin{equation}
\sum_{k\neq n}\rightarrow \int dE_k \Omega(E_k) R(\Omega |E_n-E_k|),
\label{Ek}
\end{equation}
where $\Omega$ is the average level--density and $R$ is two--level 
correlation function giving a probability density to find level with
energy $E_k$ in the interval $[E_k-dE_k,E_k+dE_k]$ at the average 
distance $|E_n-E_k|$ from the given level with energy $E_n$. Moreover, 
we believe that the energy spacings rapidly fluctuate with time so 
that they are decorrelate over time intervals of the physical interest,
\begin{eqnarray}
\overline{(E_n-E_k)(t)(E_n-E_k)(t')}=
\left\{
\begin{array}{ll}
\overline{(E_n-E_k)^2(t)},~~~t'=t
\nonumber\\
0,~~~t'\neq t
\end{array}
\right.
\label{eett}
\end{eqnarray}
Performing the ensemble averaging of Eq.~(\ref{an2ap}), one get
\begin{eqnarray}
\frac{d\overline{|a|^2}(E,t)}{dt}=2\dot{X}(t)\int_0^t dt' \dot{X}(t')
\int_{-\infty}^{+\infty} de \Omega(E-e)R(\Omega |e|)
\frac{Re(\overline{{\cal M}_{nk}(q){\cal M}_{nk}^*(X')})}{e^2}
\nonumber\\
cos(e/\hbar[t-t'])
\{\overline{|a|^2}(E-e,t')-\overline{|a|^2}(E,t')\},
\label{an2e}
\end{eqnarray}
where $e\equiv \overline{E_n-E_k}$ is a spacing between 
two energy levels and $E\equiv\overline{E_n}$ measures excitation 
of the system.

Our second step in the ensemble averaging procedure is an averaging 
over the off--diagonal matrix elements ${\cal M}_{nk}$. ${\cal M}_{nk}$ are
treated as complex random numbers with real and imaginary parts
independently Gaussian distributed, and with~\cite{wil3}
\begin{equation}
\overline{{\cal M}_{nk}(X){\cal M}_{n'k'}^*(X')}=
\overline{|{\cal M}_{nk}|^2}(\overline{E_n},\overline{E_k},X)
C(q-q')\delta_{nn'}\delta_{kk'},
\label{CXX}
\end{equation}
where $C(0)=1$ and the function $C(X-X')$ is characterized by a 
correlation length $\xi_q$ over which the matrix elements 
correlate with each other significantly for the different 
values of the external parameter $X$. To specify an energy 
distribution of the ensemble averaged squared matrix elements 
$\overline{|{\cal M}_{nk}|^2}(\overline{E_n},\overline{E_k})$, 
We take it in a quite general form~\cite{ra,kar}
\begin{equation}
\overline{|{\cal M}_{nk}|^2}(\overline{E_n},\overline{E_k},X)=
\frac{\sigma^2}{\sqrt{\Omega(\overline{E_n})\Omega(\overline{E_k})}\Gamma}
f(|\overline{E_n-E_k}|/\Gamma),
\label{f}
\end{equation}
where $\sigma^2$ is the strength and $\Gamma$ is the width
of the energy distribution of the ensemble averaged squared
matrix elements $\overline{|{\cal M}_{nk}|^2}$. Here it is 
implied that the shape of the distribution, $f$, is a decaying 
function of the energy distance between states $|\overline{E_n-E_k}|$. 

The parameter $\Gamma$ is a width of the energy distribution
$f$ and measures how strong different eigenstates are coupled 
by the transition operator $\partial\hat{H}/\partial X$. On
the other hand, $\Gamma$ determines an effective number of
states, $N\sim \Omega(E_n)\Gamma$, over which the initially 
occupied state $n$ spreads out.

Substituting Eqs.~(\ref{Cqq}) and (\ref{f}) into Eq.~(\ref{an2}),
we obtain 
\begin{eqnarray}
\frac{d\overline{|a|^2}(E,t)}{dt}=\frac{2\dot{X}(t)}{\sqrt{\Omega(E)}\Gamma}
\int_0^t dt' \dot{X}(t')\int_{-\infty}^{+\infty} de\sqrt{\Omega(E-e)} R(\Omega |e|)
f(|e|/\Gamma)C(X-X')
\nonumber\\
\frac{cos(e/\hbar[t-t'])}{e^2}
\{\overline{|a|^2}(E-e,t')-\overline{|a|^2}(E,t')\},
\label{an2em}
\end{eqnarray}

\section{Different regimes of the drivn quantum dynamics}
\label{dr}

Assuming that the occupancy of the given state with energy $E$ 
changes mainly due to the direct interlevel transitions from 
the close--lying states located at the distances $|e|<<E$, we
enable to truncate the following expansion, 
\begin{eqnarray}
\sqrt{\Omega (E-e)}\{\overline{|a|^2}(E-e,t')-\overline{|a|^2}(E,t')\}
=-\sqrt{\Omega (E)}\frac{\partial \overline{|a|^2}(E,t')}{\partial E}e  \nonumber \\
+\frac{1}{2\sqrt{\Omega (E)}}\frac{d\Omega (E)}{dE}\frac{\partial \overline{|a|^2}(E,t')}
{\partial E}e^{2}+\frac{\sqrt{\Omega (E)}}{2}\frac{\partial ^{2}
\overline{|a|^2}(E,t')}{\partial E^{2}}e^{2}+(...)e^3 + O(e^4)
\label{expan}
\end{eqnarray}
to $e^3$--order terms.

The expansion~(\ref{expan}) leads us to a non--Markovian
Fokker--Planck equation for the ensemble averaged occupancy 
$\overline{\rho}(E,t)$ of the given quantum state with the energy $E$,
\begin{equation}
\Omega(E)\frac{\partial \overline{|a|^2}(E,t)}{\partial t}\approx \sigma^{2}
\dot{X}(t)\int_{0}^{t}dt'\dot{X}(t')C(X[t]-X[t']) K(t-t')\frac{\partial }
{\partial E}\bigg[\Omega (E)\frac{\partial \overline{|a|^2}(E,t')}{\partial E}\bigg],
\label{Pnonm}
\end{equation}
where
\begin{equation}
K(t-t')=\frac{1}{\Gamma}Re\bigg(\int_{\infty}^{+\infty}
f(|e|/\Gamma)R(\Omega |e|)exp(\frac{ie[t-t']}{\hbar})de\bigg).
\label{K}
\end{equation}
Eq.~(\ref{Pnonm}) can be understood in a probabilistic sense as
an dynamical equation for a probability distribution function 
$P(E,t)\equiv \overline{|a|^2}(E,t)\Omega(E)$ showing the 
relative number of quantum states
with energies which lie in the interval $[E,E+dE]$. From this
point of view, we can speak about quantum mechanical diffusion
of energy caused by the direct interlevel transitions between
energy states. Two different time scales, appearing in Eq.
(\ref{Pnonm}), determine a non--Markovian character of the
energy diffusion. The first one, $\tau_{\xi} \sim \xi/\dot{X}$,
originates from the correlations between the ensemble averaged 
squared matrix elements~(\ref{CXX}) existing at different values 
of the external time--dependent parameter $X[t]$. The second one, 
$\tau_{\Gamma} \sim \hbar/\Gamma$, is defined by the energy--dependence 
of the ensemble averaged squared matrix elements~(\ref{f}).

To study how these time scales define the quantum diffusive 
dynamics~(\ref{Pnonm}), we use a number of simplifying
assumptions. First of all, we shall consider quantum systems
with constant average level--density, $\Omega(E)=\Omega_0$,
driven with a constant velocity, $X[t]=V_0 \cdot t$. Secondly, 
we take the correlation function $C(X-X')$ (\ref{CXX}) and the 
memory kernel $K(t-t')$ (\ref{K}) in a simple exponential form,
\begin{equation}
C(X-X')=exp\left(-\frac{|X-X'|}{\xi}\right)
\label{Cexp}
\end{equation}
and
\begin{equation}
K(t-t')=K_0 \cdot exp\left(-\frac{|t-t'|}{\hbar/\Gamma}\right),
\label{Kexp}
\end{equation}
where $K_0$ is some constant independent of the width $\Gamma$.
Thus, we obtain a non--Markovian diffusion equation of the form
\begin{equation}
\frac{\partial P(E,t)}{\partial t}=\sigma^2 K_0 V_0^2
\int_0^t exp\left(-\frac{|t-t'|}{\tau}\right)
\frac{\partial^2 P(E,t')}{\partial E^2}dt',
\label{Pnonm1}
\end{equation}
with the normalization condition 
\begin{equation}
\int P(E,t)dE=1,
\label{Pnorm}
\end{equation}
and the initial condition
\begin{equation}
P(E,t=0)=\delta(E-E_0),
\label{P0}
\end{equation}
where $E_0$ is the initial excitation energy of the system.
Here the different time scales, $\tau_{\xi}=\xi/V_0$, caused
by the $X$--correlations of the ensemble averaged squared
matrix elements~(\ref{CXX}), and $\tau_{\Gamma}=\hbar/\Gamma$,
due to the energy--dependence of the squared matrix elements
(\ref{f}), appear in Eq.~(\ref{Pnonm1}) in the following
combination
\begin{equation}
\frac{1}{\tau}=\frac{1}{\xi/V_0}+\frac{1}{\hbar/\Gamma}.
\label{tau}
\end{equation}
In fact, a parameter $\tau$ measure the strength of the memory 
effects in the energy diffusion~(\ref{Pnonm1}) and from that
perspective, it is relevant to call it a memory time of the
quantum diffusion dynamics. Depending on that parameter, the
different regimes of the quantum diffusion dynamics~(\ref{Pnonm1}) 
can be distingwuished. To show this, we differentiate over
time both sides of Eq.~(\ref{Pnonm1}) and reduce it to the
second order in time differential equation
\begin{equation}
\frac{\partial^2 P}{\partial t^2}+
\frac{1}{\tau}\frac{\partial P}{\partial t}=
\sigma^2 K_0 V_0^2 \frac{\partial P}{\partial E^2}.
\label{Pnonm2}
\end{equation}

\subsection{Diffusion regime (weak memory effects)}
\label{nd}

$\tau\rightarrow 0$. This is a limit of extremely
small values of the memory time $\tau$, when it is the shortest
time scale of the system and the memory effects in the
system's dynamics are of minor role. By neglecting the first
term in the left--hand side of Eq.~(\ref{Pnonm2}) compared
to the second one, we end up with a normal diffusion
regime of the quantum dynamics~(\ref{Pnonm1}), 
\begin{equation}
\frac{\partial P}{\partial t}=
\sigma^2 K_0 V_0^2 \tau\frac{\partial P}{\partial E^2}.
\label{Pnonmnd}
\end{equation}
Therefore, we can claim that in the case of the weak memory effects
in the quantum driven dynamics~(\ref{schr}) we have the normal 
time diffusion of the occupancies of adiabatic states when 
a variance of its energy distribution, 
$v^2_E=\int E^2 P(E,t)dE-(\int E P(E,t)dE)^2$, behaives linearly
with time,
\begin{equation}
v^2_E=\frac{\hbar \sigma^2 K_0 V_0^2 \xi}{\hbar V_0+\xi\Gamma} \cdot t,
\label{v2End}
\end{equation}
see Eq.~(\ref{tau}).
It is interesting that the relationship between the time scales
$\tau_{\xi}=\xi/V_0$ and $\tau_{\Gamma}=\hbar/\Gamma$ leads to
a principly different behaviour of the energy diffusion $v^2_E$
as a function of the driven velocity $V_0$. Let us consider two
limiting cases:

(i) $\tau_{\Gamma}<<\tau_{\xi}~~~(\hbar /\Gamma<<\xi/V_0)$.
This situation is realized at either semiclassical limit
($\hbar\rightarrow 0$) or fairly large widths $\Gamma$ of
the energy--distribution~(\ref{f}) of the ensemble averaged
squared matrix elements. In this case the energy variance
(\ref{v2End}) behaives with the driven velocity $V_0$ quadratically,
\begin{equation}
v^2_E\sim \frac{\hbar\sigma^2 K_0}{\Gamma}\cdot V_0^2.
\label{v2End1}
\end{equation}
It should be stressed that the energy diffusion drops out with
the growth of the width $\Gamma$. This feature can be understood
as follows. The width $\Gamma$ defines an effective number of states 
$N\sim\Gamma\Omega_0$ coupled
by the transition operator $\partial\hat{H}/\partial X$ at the given
excitation $E$. The initially occupied many body state with energy 
$E$ will spread out over $N$ neighboring states, resulting in a gradual 
equilibration of the driven quantum system~(\ref{schr}). The larger 
$\Gamma$, the closer the quantum system to the equilibrium and 
therefore, the weaker the energy diffusion. Also note that in the
limit of large widths $\Gamma$, the energy diffusion is independent
on the correlation length $\xi$ (\ref{CXX}) of the distribution
of the ensemble averaged squared matrix elements.

\bs

(ii) $\tau_{\xi}<<\tau_{\Gamma}~~~(\xi/V_0<<\hbar/\Gamma)$. This
condition is reached at quite large driven velocities $V_0$ or at
relatively small values of the correlation length $\xi$. In that
case we obtain a significant suppression of the energy diffusion
when the energy variance $v^2_E$ becomes linearly proportional
to the driven velocity,
\begin{equation}
v^2_E\sim \xi \sigma^2 K_0 \cdot V_0.
\label{v2End2}
\end{equation}
We see that the energy diffusion linearly grows with the increase
of the $X$--correlations~(\ref{CXX}) between ensemble averaged
squared matrix elements.

\subsection{Ballistic regime (strong memory effects)}
\label{br}

$\tau\rightarrow \infty$. This is opposite limiting situation 
of the very strong memory effects when the memory time
$\tau$ is assumed to be larger than the time of physical
interest. Now the second term in the left--hand side of
Eq.~(\ref{Pnonm2}) is quite small and we come to a
telegraph--like equation
\begin{equation}
\frac{\partial^2 P}{\partial t^2}=
\sigma^2 K_0 V_0^2 \frac{\partial P}{\partial E^2},
\label{Pnonmbr}
\end{equation}
whose solution is given by the sum of two delta peaks,
\begin{equation}
P(E,t)=\frac{1}{2}\left[\delta(E-E_0-\sigma^2 K_0 V_0^2\cdot t)
+\delta(E-E_0+\sigma^2 K_0 V_0^2\cdot t)\right].
\label{Pbr}
\end{equation}
Here we have a ballistic regime of the quantum dynamics
(\ref{Pnonm1}) when the energy variance quadratically 
depends on time
\begin{equation}
v^2_E=\sigma^2 K_0 V_0^2\cdot t^2.
\label{v2Ebr}
\end{equation}
Microscopically speaking the ballistic regime~(\ref{Pnonmbr})
corresponds to the situation when the initial distribution of 
the adiabatic state with energy $E_0$ splites in two equal
delta--pulses propagating in the energy space in opposite directions 
with the constant velocity $\sigma \sqrt{K_0} V_0$.

In the intermediate situation of finite--sized memory effects 
the parameter $\tau$ in Eq.~(\ref{Pnonm2}) plays a role of
a crossover time, i. e., when $\tau$ separates a short--time
regime, $t<\tau$ of the ballistic propagation of the probability
between different states~(\ref{Pnonmbr}) from a long--time regime, 
$t>\tau$, of the normal diffusive behaviour of the occupancies of 
the adiabatic states~(\ref{Pnonmnd}).

\section{Summary}
\label{sum}

In the paper we have addressed the general problem of the response
of complex quantum systems on a parametric external driving
represented by a single time--dependent classical variable $X[t]$. 
Driven dynamics of the quantum system has been started to discuss 
in the adiabatic basis of the eigen--energies and eigen--functions 
of the system's Hamiltonian $\hat{H}[X]$ (\ref{adiab}) found at each 
fixed value of the external parameter $X$. Have considering the 
perturbative response~(\ref{alpha}) of the system, we have obtained 
a closed set of equations~(\ref{an2ap}) determining the time evolution 
of the occupancies of adiabatic states. The obtained equations has a 
form of the rate equation for occupancy of given quantum state $n$ 
which varies with time due to direct interlevel transitions from all 
other states $m$. 

Then, we have applied the random matrix theory to study the
driven dynamics~(\ref{an2ap}). Thus, we have performed ensemble
averaging over spacings between energy levels and off--diagonal
matrix elements $(\partial \hat{H}/\partial X)_{nm}$. The latter
has been modelled by independent Gaussian distributed random variables
(\ref{CXX}) where we take into account both possible time--correlations
of the coupling matrix elements $(\partial \hat{H}[X]/\partial X)_{nm}$
and its energy--dependence~(\ref{f}). The correlation function $C(X-X')$, 
defining how strong the ensemble averaged matrix elements correlate
at different values $X$ and $X'$ of the external parameter, has been
characterized by a correlation length $\xi$. The energy distribution
of the ensemble averaged squared matrix elements has been described
by a width $\Gamma$ that determines an effective number of states
over which the initially occupied state will spread out due to the
external parametric driving. Thus, five parameters enter into our
model of the driven quantum dynamics~(\ref{an2em}): the velocity
of driving $\dot{X}$, the strength of coupling of the quantum
system to the externel parameter $\sigma^2$ (\ref{f}), the average
density of states of the system $\Omega(E)$ at given energy $E$ 
and the parameters of the matrix elements' distribution $\xi$
and $\Gamma$. In the sequel, we have studied the response of 
a quantum system with constant level--density $\Omega(E)=\Omega_0$
on an external driving with constant velocity $\dot{X}[t]=V_0$.

In that case we are able to describe the quantum dynamics in terms
of non--Markovian Fokker--Planck equation~(\ref{Pnonm1}) for the probability
distribution function $P(E,t)$ giving a relative number of quantum states
with energies in the interval $[E,E+dE]$ at time $t$. The non--Markovian
character of the quantum dynamics~(\ref{Pnonm1}) is defined by time
$\tau$ which is a geometric average~(\ref{tau}) of two time scales,
$\tau_{\xi}=\xi/V_0$, caused by $X$--correlations between the ensemble
averaged matrix elements, and $\tau_{\Gamma}=\hbar/\Gamma$, appearing
due to the energy--dependence of the matrix elements. We have analyzed
the energy diffusion~(\ref{Pnonm1}) in two limiting cases of extremely
small and large memory times $\tau$. In the first case, we obtain a
normal diffusion regime of the quantum dynamics~(\ref{Pnonmnd}) with
the energy variance $v^2_E\sim \hbar\sigma^2 V_0^2/(V_0/\xi+\Gamma/\hbar)\cdot t$.
This regime corresponds to the Markovian limit of the driven quantum
dynamics when the memory kernel of non--Markovian Fokker--Planck equation
(\ref{Pnonm1}) becomes sharply peaked function of time and can be
effectively modelled by a delta--function.
We see that in this regime the energy diffusion drops out with the
width $\Gamma$ of the energy--distribution of the ensemble averaged
squared matrix elements~(\ref{f}). In fact, that means that as far as
we increase the width $\Gamma$ the initially occupied many body state 
will distribute to large number of neighboring states which in turn
results in a faster equilibration of the system. Threfore, the larger 
$\Gamma$ the weaker the energy diffusion. It is also important that
the energy diffusion can be strongly suppressed
depending on the correlation length $\xi$. If $\xi$ is quite big then
the energy variance behaves quadratically with the driving velocity,
$v^2_E \sim V_0^2$, while at relatively small values of the correlation
length $\xi$ the energy variance is significantly reduced and
becomes linear proportional to the velocity $v^2_E \sim \xi\cdot V_0$.
This feature is quite natural because any correlations in the sytem
give rise to more regular dynamics and as a consequence of that,
by decreasing the size of correlations (decreasing the correlation
length $\xi$) we make dynamics more chaotic (more diffusive). 

In the other Markovian limit, reached at fairly large values of the 
memory time, $\tau\rightarrow \infty$, we get ballistic regime 
of the quantum dynamics~(\ref{Pnonmbr}) when the variance of energy 
behaives quadratically with time~(\ref{v2Ebr}). Here the memory kernel 
in Eq.~(\ref{Pnonm1}) can be well approximated by one and the 
non--Markovian Fokker--Planck equation~(\ref{Pnonm1}) can be 
reduced to the telegraph--like equation which is of second order 
in time. Now we have principly different picture of the driven 
quantum dynamics: the initial distribution of the occupied many 
body state splites into two equal pulses which then begin to move 
from each other in energy space with constant velocity 
$\sim \sigma V_0$. For moderate values of the memory time $\tau$,
we expect that at short times $t<\tau$ the ballistic regime
of the quantum dynamics~(\ref{Pnonmbr}) is observed while at
large times $t>\tau$ we have the normal diffusion~(\ref{Pnonmnd}) 
of the occupancies of adiabatic states. 

Of course, the first natural continuation of the present study is
to consider the non--perturbative response of a complx quantum
system, i. e. when the parameter~(\ref{alpha}) of our perturbation 
expansion~(\ref{formal}) is not small and we have to include all 
terms in that expansion. Secondly, it is of big interest to study 
the macroscopic manifestation of found different regimes of the 
quantum dynamics, i. e. when the parameter $X[t]$ is not a tunable 
parameter but it is rather some effective coordinate like collective 
deformation parameters of nuclear or atomic physics.

\section{Acknowledgements}

The work of S.V.R. on the project Nuclear collective dynamics for high
temperatures and neutron-proton asymmetries was supported (partially) 
by the Program Fundamental research in high energy physics and nuclear physics (international collaboration) at the Department of Nuclear Physics and Energy 
of the National Academy of Sciences of Ukraine.

\end{document}